\documentclass[preprint2]{aastex}

\shorttitle{Orbit Evolution of Algol Binaries with a CB Disk}
\shortauthors{Chen, Li, \& Qian}

\begin{document}


\title{Orbital Evolution of Algol Binaries with a Circumbinary Disk}


\author{Wen-Cong Chen, Xiang-Dong Li}
\affil{Department of Astronomy, Nanjing University,
    Nanjing 210093, China}
    \and
\author{Sheng-Bang Qian}
\affil{National Astronomical Observatories/Yunnan Astronomical
Observatory, Chinese Academy of Sciences, P.O. Box 110, Kunming
650011,  China}



\begin{abstract}
It is generally thought that conservative mass transfer in Algol
binaries causes their orbits to be wider, in which the less
massive star overflows its Roche-lobe. The observed decrease in
the orbital periods of some Algol binaries suggests orbital
angular momentum loss during the binary evolution, and the
magnetic braking mechanism is often invoked to explain the
observed orbital shrinkage. Here we suggest an alternative
explanation, assuming that a small fraction of the transferred
mass forms a circumbinary disk, which extracts orbital angular
momentum from the binary through tidal torques. We also perform
numerical calculations of the evolution of Algol binaries with
typical initial masses and orbital periods. The results indicate
that, for reasonable input parameters, the circumbinary disk can
significantly influence the orbital evolution, and cause the orbit
to shrink on a sufficiently long timescale. Rapid mass transfer in
Algol binaries with low mass ratios can also be accounted for in
this scenario.
\end{abstract}


\section{Introduction}

An Algol binary is a semidetached binary system consisting of (1)
an early type, main sequence primary component which does not fill
its Roche lobe, and (2) a lobe-filling, less massive star that is
substantially above the main sequence. The less massive star is
cooler, fainter and larger \citep{Giur81,pete01}. It is believed
that star (2) is initially more massive, and evolves first to
overflow its Roche lobe to transfer mass to star (1). After the
rapid mass exchange, the lobe-filling star (2) becomes less
massive in the Algol binary stage.

The evolution of Algol binaries has been investigated extensively
(e.g., Plavec 1968; Paczynski 1971; Refsdal \& Weigert 1969).
Although conservative mass transfer seems to roughly reproduce the
observed characteristics of a considerable fraction of Algol
binaries (e.g. Nelson \& Eggleton 2001; De Loore \& van Rensbergen
2004), nonconservative evolution with mass and orbital angular
momentum loss taken into account is required for better comparison
between theories and observations (Thomas 1977; Refsdal et al.
1974; Sarna 1993). With regard to the orbital evolution,
conservative mass transfer in Algol binaries leads to increase of
the orbital periods because mass is transferred from the less
massive star to the more massive component (Huang 1963). However,
there is strong evidence showing that the orbital periods of some
Algol binaries are actually decreasing
\citep{qian00a,qian00b,qian01a,qian01b,qian01c,qian02,lloy02}.
Although it is uncertain whether the measured period changes are
long-term (secular) changes or transient fluctuations, they
clearly demonstrate that these systems must undergo mass and
angular momentum loss during the evolution. Moreover,
\citet{qian02} find that the decreasing rates $\dot{P}_{\rm orb}$
scale with the orbital periods $P_{\rm orb}$ roughly as
$-\dot{P}_{\rm orb}\propto P_{\rm orb}$.

This orbital decay may be explained by angular-momentum loss via
magnetic stellar winds \citep{verb81}. Sarna, Muslimov, \& Yerli
(1997) and Sarna, Yerli, \& Muslimov (1998) proposed that dynamo
action can occur in mass-losing stars in Algol-type binaries, and
produce large-scale magnetic fields, which lead to magnetic
braking of the stars. However, there could exist some potential
difficulties for the magnetic braking mechanism (see below). Here
we suggest an alternative interpretation assuming that a small
fraction of the transferred mass would form a circumbinary (CB)
disk surrounding the binary system \citep[see also][]{van94}.
Previous works \citep{taam01,chen06} have already indicated that
the CB disk can efficiently remove the orbital angular momentum
and accelerate the binary evolution in accreting white dwarf or
neutron star binaries. We describe the input physics that is
necessary for the evolution model in section 2. Numerically
calculated results for the evolutionary sequences of Algol
binaries in two cases are presented in section 3. We make brief
discussion and conclude in section 4.

\section{Input Physics}

We consider an Algol binary consisting of a lobe-filling, less
massive star (of mass $M_{2}$), and an early type, more massive
companion (of mass $M_{1}$). The total orbital angular momentum of
the system in a circular orbit is
\begin{equation}
J=a^{2}\mu\frac{2\pi}{P_{\rm orb}},
\end{equation}
where $a=(GM_{\rm T}P_{\rm orb}^{2}/4\pi^{2})^{1/3}$ is the binary
separation and $\mu=M_{1}M_{2}/M_{T}$ is the reduced mass of the
binary system, $G$, and  $M_{\rm T}$ are the gravitational
constant and the total mass of system, respectively. We neglect
the spin angular momentum of the components because it is smaller
compared with the total orbital angular momentum of the system.
Similar to \citet{rapp83}, in the calculations we consider two
types of mechanisms for orbital angular momentum loss from the
binary system, which are described as follows.

\subsection{Magnetic Braking}

\citet{verb81} point out that single low-mass main-sequence stars
will undergo braking by the coupling between the magnetic field
and the stellar winds. The magnetic braking mechanism assumes that
the ionized particles stream out along the magnetic field lines
which come off the convection envelope of the stars. The loss of
the specific angular momentum in the stellar winds is very large
because the outflowing material is tied in the magnetic field
lines to corotate with the stars out to a long distance ($\ga 5
-10$ stellar radii) \citep{scha62}. In an Algol binary, the
angular momentum loss of the evolved star by magnetic braking
cause it to spin down. However, the tidal forces of the system
will continuously act to spin the evolved star up back into
corotation with the orbital rotation. The spin up takes place with
the expense of the orbital angular momentum. Hence magnetic
braking indirectly carries away the orbital angular momentum of
the binary system.

The mass loss due to the magnetic winds during the mass transfer
is generally very small, so we can assume that the total mass is
conserved, and the orbital evolution is connected to the angular
momentum loss rate and the mass transfer rate as follows
\begin{equation}
\frac{\dot{P}_{\rm orb}}{P_{\rm orb}}=\frac{3\dot{J}_{\rm
MB}}{J}-\frac{3\dot{M}_2}{M_2}(1-q),
\end{equation}
where $q=M_2/M_1$ is the mass ratio, and $\dot{J}_{\rm MB}$ is the
angular momentum loss rate due to magnetic braking,
\begin{equation}
\dot{J}_{\rm MB}\simeq -3.8\times
10^{-30}M_2R_{\odot}^4(R_2/R_{\odot})^{\gamma}\omega^3\,{\rm
dyn\,cm},
\end{equation}
where $\gamma$ is a dimensionless parameter in the range from 0 to
4, and $\omega$ is the angular velocity of the star (Rappaport et
al. 1983). Since $q<1$ and $\dot{M}_2<0$, the second term on the
right hand side of Eq.~(2) always causes the orbit to increase.
Orbital shrinkage occurs only when the magnitude of the first term
is larger than that of the second one, i.e.,
\begin{equation}
\frac{\dot{P}_{\rm orb}}{P_{\rm orb}}\sim \frac{3\dot{J}_{\rm
MB}}{J}.
\end{equation}
One can show that with the magnetic braking law of Rappaport et
al. (1983), the orbital decay rates with Eq.~(4) are several
orders lower than the observed ones. Moreover, from Eqs.~(3) and
(4) we can derive the following relation,
\begin{equation}
-\dot{P}_{\rm orb}\propto P_{\rm orb}^{(2\gamma-7)/3},
\end{equation}
where we have used the relation for the radius of the lobe-filling
star $R_2=0.462[q/(1+q)]^{1/3}a$ with $0.1\la q\la 0.8$ (Paczynski
1971). As we can see, only when $\gamma\ga 4$ can we have a
positive correlation between $-\dot{P}_{\rm orb}$ and $P_{\rm
orb}$. More recent studies based on observations of rapidly
rotating stars show that the above magnetic braking form may be
inadequate \citep{quel98,sill00,andr03}. \citet{sill00} instead
propose an empirical angular momentum loss prescription
\begin{eqnarray}
\dot{J}_{\rm MB}=-K_{w}\omega
^{3}\left(\frac{r_2}{m_2}\right)^{1/2},\quad \quad \quad
\omega\leq\omega_{\rm crit} \nonumber\\
 \dot{J}_{\rm
MB}=-K_{w}\omega \omega_{\rm
crit}^{2}\left(\frac{r_2}{m_2}\right)^{1/2},\quad \quad
\omega>\omega_{\rm crit}
\end{eqnarray}
where $K=2.7\times 10^{47} \rm g cm^{2}$, $\omega_{\rm crit}$ is
the critical angular velocity at which the angular momentum loss
rate reaches a saturated state, $r_2$ and $m_2$ are $R_2$ and
$M_2$ in solar units, respectively. \citet{kim96} suggest that
$\omega_{\rm crit}$ is inversely proportional to the convective
turbulent timescale in the star at 200 Myr age,
\begin{equation}
\omega_{\rm crit}=\omega_{\rm
crit,\odot}\frac{\tau_{\odot}}{\tau}.
\end{equation}
The magnitude of $\dot{J}_{\rm MB}$ under this magnetic braking
law is at least one order of magnitude smaller than that in the
Rappaport et al. (1983) model with $\gamma=4$ (Andronov et al.
2003), suggesting that more efficient angular momentum loss
mechanism may be required to account for the orbit decay in Algol
binaries.

\begin{figure}
\epsscale{.80} \plotone{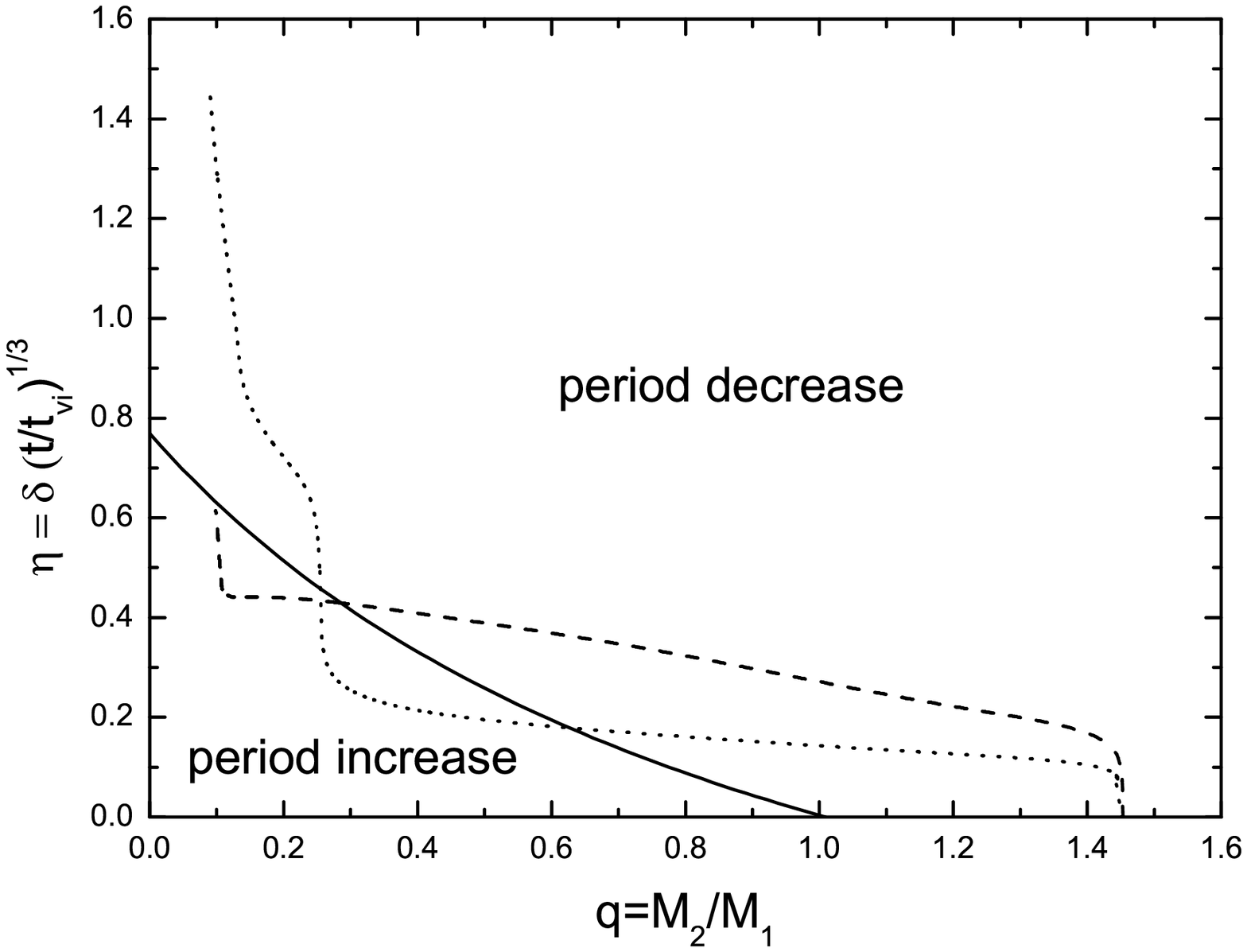} \caption{ The figure of $\eta$
against the mass ratio $q$. The solid curve denote the division
between period increase and decrease. The dashed and dotted curves
correspond to the numerical results of Algol binary evolution with
the initial orbital periods $P_{\rm orb,i} =3.4$ d
$(\delta=0.032)$ and 1.4 d $(\delta=0.01)$ and initial masses
$(M_2, M_1)=(3.0, 2.0) M_{\odot}$, respectively.\label{fig1}}
\end{figure}

\subsection{CB Disk}
Once Roche lobe overflow occurs in semidetached binaries, some
fraction of the transferred material may leave the system in
various ways. The mass outflow in cataclysmic variables (CVs) was
noticed by \citet{warn87} from the studies of UV and optical
resonance lines.

Mass loss can also lead to orbital decrease in mass exchanging
binaries. We first consider the case that a fraction $\delta$ of
the transferred matter is lost from the binary system, carrying
away the specific angular momentum of the secondary star
\citep{rapp83}. The change rate of orbital period can be written
as
\begin{equation}
\frac{\dot{P}}{P}=-3\frac{\dot{M}_{2}}{M_{2}}
\left[1-q(1-\delta)-\frac{\delta}{1+q}-\frac{1}{3}\frac{\delta
q}{1+q}\right].
\end{equation}
To account for the secular decreasing of orbital period, the mass
loss fraction $\delta$ from the binary system should satisfy
\begin{equation}
\delta > \frac{3(1-q^{2})}{3-2q-3q^{2}}.
\end{equation}
For $q=0.5$, it requires $\delta>1.8$. Apparently this is
impossible. Hence angular momentum loss via simple mass loss
cannot interpret the orbital shrinking of some Algol binaries with
low mass ratio.

In a theoretical view, \citet{spru94} proposed that part of the
outflow may be in the form of a slow wind near the orbital plane.
It has been argued that during mass exchange the lost matter may
form a disk structure surrounding the binary system rather than
leave the binary system \citep{van73,van94}. More recently,
\citet{taam01} and \citet{spru01} suggested that a Keplerian CB
disk could be formed as a result of mass outflow in CVs. The CB
disk can extract the orbital angular momentum from th binary orbit
through tidal torques.

Since the mass transfer processes are similar in CVs as in Algol
binaries, we assume that a constant fraction $\delta$ of the
transferred mass feeds into the CB disk surrounding the Algol
system, and $\dot{M}_{1}=-(1-\delta)\dot{M}_{2}$ (if the mass
transfer rate is sub-Eddington). Same as in the standard thin
accretion disks, the viscous torque in the CB disk is
\citep{Shak73}
\begin{equation}
T=-2\pi r\nu\Sigma r\frac{d\Omega}{dr}r,
\end{equation}
where $\nu$ is the viscosity, $r$ the radius, $\Sigma$ the surface
density, and $\Omega=[G(M_1+M_2)/r^{3}]^{1/2}$ the local orbital
angular velocity in the disk.

Assume that the gravitational interaction of the binary with the
disk occurs locally at the inner edge $r_{\rm i}$ of the
disk\footnote{In the following we use the subscript i to denote
quantities evaluated at $r_{\rm i}$}, angular momentum from the
binary feeds into the disk at a rate proportional to the surface
density $\Sigma_{\rm i}$. The viscous torque exerted by the CB
disk on the binary can be shown to be \citep{taam01,spru01}
\begin{equation}
T_{\rm i}\equiv\dot{J}_{\rm CB}=\gamma\left(\frac{2\pi a^2}{P_{\rm
orb}}\right)\delta\dot{M}_{2}\left(\frac{t}{t_{\rm
vi}}\right)^{p},
\end{equation}
where $\gamma^{2}=r_{\rm i}/a$, $t$ is the mass transfer time, and
$p$ is a parameter determined by the viscosity. Suppose $\nu$ is
the function of $r$ and $\Sigma$ with the following form
\citep{spru01}
\begin{equation}
\nu=\nu_{\rm i}\left(\frac{r}{r_{\rm
i}}\right)^{n}\left(\frac{\Sigma}{\Sigma_{\rm i}}\right)^{m},
\end{equation}
then $p=(m+1)/[2(2m+2-n)]$. In the following calculations we set
$m=n=1$, so $p=1/3$.

By simple algebra, from the above equations we obtain the angular
momentum loss rate via the CB disk
\begin{equation}
\dot{J}_{\rm CB}=\delta\left(\frac{t}{t_{\rm
vi}}\right)^{1/3}\dot{M_{2}}j_{\rm CB},
\end{equation}
where
\begin{equation}
j_{\rm CB}=\gamma\frac{J}{\mu}.
\end{equation}
is the specific orbital angular momentum of the disk material at
$r_{\rm i}$.

The viscous timescale $t_{\rm vi}$ at the inner edge of disk is
\begin{equation}
t_{\rm vi}=\frac{4r_{\rm i}^{2}}{3\nu_{\rm i}},
\end{equation}
and $\nu_{\rm i}$ is given with the standard $\alpha$ prescription
\citep{Shak73}
\begin{equation}
\nu_{\rm i}=\alpha_{\rm SS}c_{\rm s} H_{\rm i},
\end{equation}
where $\alpha_{\rm SS}$, $c_{\rm s}$, $H_{\rm i}$ are the
viscosity parameter (in the following calculations we set
$\alpha_{\rm SS}=0.01$), the sound speed, and the scale height of
disk at $r_{\rm i}$, respectively. Assuming that the disk is
hydrostatically supported and geometrically thin with $H_{\rm
i}/r_{\rm i}\sim 0.03$ \citep{bell04}, we have
\begin{equation}
c_{\rm s}\simeq\Omega_{\rm K} H_{\rm i}.
\end{equation}

If we consider angular momentum loss only via the CB disk, the
changing rate of the orbital period is then
\begin{equation}
\frac{\dot{P}}{P}=-3\frac{\dot{M}_{2}}{M_{2}}
\left[1-\gamma\eta(1+q)-q(1-\delta)-\frac{1}{3}\frac{\delta
q}{1+q}\right],
\end{equation}
where $\eta=\delta(t/t_{\rm vi})^{1/3}$. Since $\dot{M}_{2}<0$,
the orbital period will decrease when
\begin{equation}
\eta > \frac{1-q(1-\delta)-(1/3)\delta q/(1+q)}{\gamma(1+q)}.
\end{equation}
For conservative evolution, $\delta=\eta=0$, Eq.~(19) recovers to
$q>1$ as expected.

In Fig.~1 we plot $\eta$ against $q$  in the solid curve when the
orbital period is constant and $\delta\ll 1$. Here we take $r_{\rm
i}/a=\gamma^{2}=1.7$ \citep{arty94}. Figure 1 shows that the CB
disk presents a plausible mechanism removing the orbital angular
momentum to explain the decrease of the orbital periods for
reasonable values of $\eta$ (or $\delta$). Generally a larger
$\eta$ is required for a smaller $q$. The dotted and dashed lines
show two examples of Algol binary evolution, revealing how the
orbital period evolution depends on the initial parameters. As
seen from Eq.~(14), the larger orbit, the larger angular momentum
loss rate via the CB disk. This naturally results in a positive
correlation between $\dot{P}_{\rm orb}$ and $P_{\rm orb}$.


\section{Numerical Results}

In this section we present the results of numerical calculations
of the Algol binary evolutions in two typical cases. We have
adopted an updated version of the evolution code developed by
\citet{eggl71,eggl72} \citep[see also][]{han94,pols95}. We
consider both the angular momentum loss mechanisms described in
section 2, adopting the modified magnetic braking law
(Eqs.~[6]-[7]) for evolved stars with mass $\la 1.5M_{\odot}$. We
take solar chemical compositions ($X=0.7$, $Y=0.28$, and $Z=0.02$)
for both stars, and the ratio of the mixing length to the pressure
scale height to be 2.0.

\noindent{\it (1) Initial masses $(M_2, M_1)=(3.0, 2.0)M_{\odot}$
and orbital period $P_{\rm orb,i}\simeq1.4$ days}

In a proto-Algol system with initial orbital period $P_{\rm
orb,i}<\sim 1-2$ days, the more massive star will fill its Roche
lobe when it is still on the main sequence. This kind of Case A
binary evolution has been investigated extensively (e.g., Eggleton
2000). Here we calculate the evolutionary sequences for a binary
with initial masses $M_{2}=3M_{\odot}$, $M_{1}=2M_{\odot}$, and
the initial orbital period $P_{\rm orb,i}\simeq 1.4$ days,
assuming that a small fraction $\delta$ ($=0.01$, 0.005) of the
transferred mass forms a CB disk surrounding the system. Figure 2
shows the mass ratio, the orbital period and the mass transfer
rate as a function of the time of mass transfer. It is clear that
the mass transfer proceeds initially on short ($\la 1$ Myr),
thermal-timescale, then on a much longer (hundreds of Myr)
nuclear-timescale when $q<1$, this is why we usually observe less
massive donor stars in Algol binaries. The orbital period first
decreases until $q\sim 0.6$. Its further evolution depends on the
effect of orbital angular momentum loss by the CB disk. When
$\delta=0.005$ the evolution is similar as the conservative
evolution with continuously increasing orbital period. When
$\delta=0.01$ the angular momentum loss is strong enough that at
$q\sim 0.25$, $P_{\rm orb}$ turns to decrease. The second orbital
decay phase occupies most of the mass transfer lifetime. In both
cases the mass transfer rates have a peak values $\sim 3\times
10^{-6}\rm  M_{\odot}yr^{-1}$ at the beginning of Roche lobe
overflow, then last a plateau phase $\sim 100-200$ Myr. It is also
noted that, in both cases the mass transfer rates reach a peak
when $t\sim 150-300$ yr, indicating that the CB disk-induced
angular momentum loss begins accelerating the mass transfer. (The
evolution of $\eta$ was plotted against the mass ratio $q$ with
the dotted curve in Fig.~1.)

\begin{figure}
\epsscale{.3} \plotone{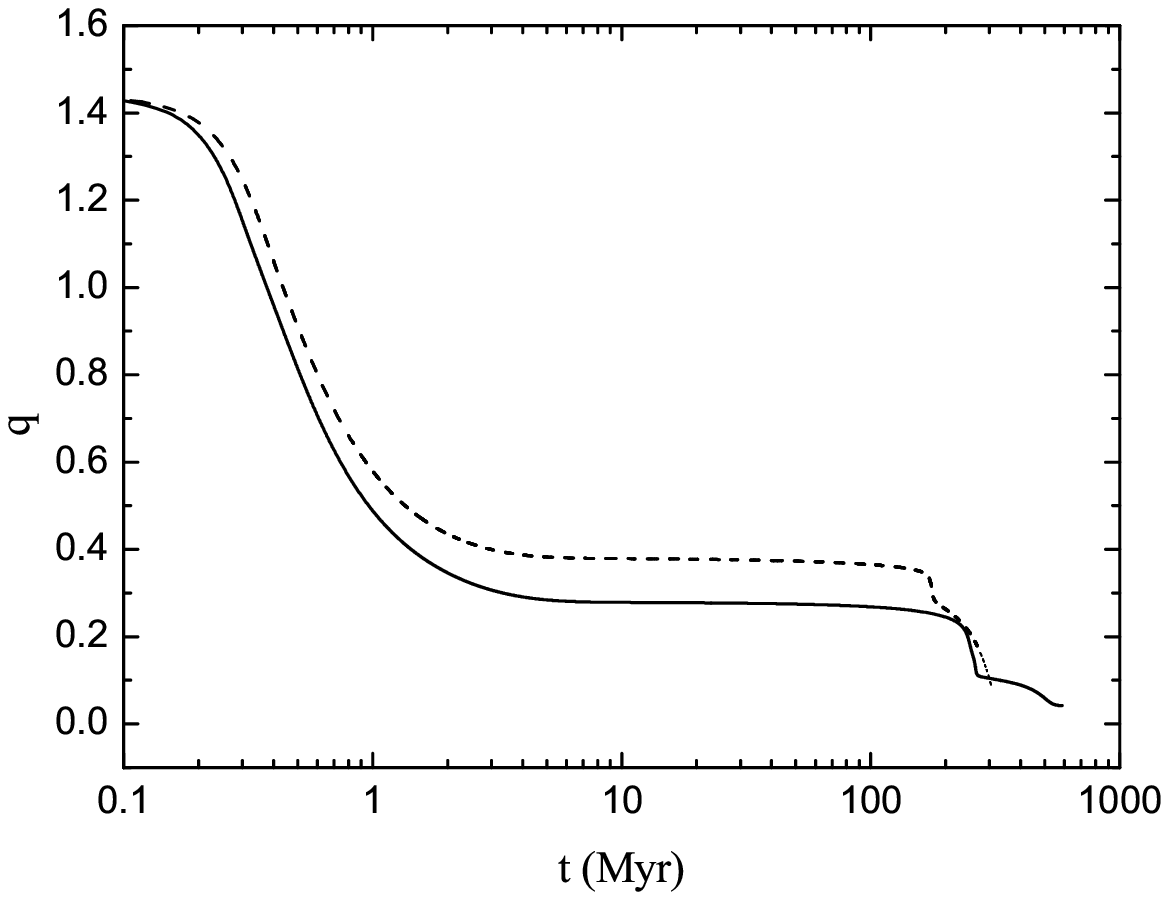}\plotone{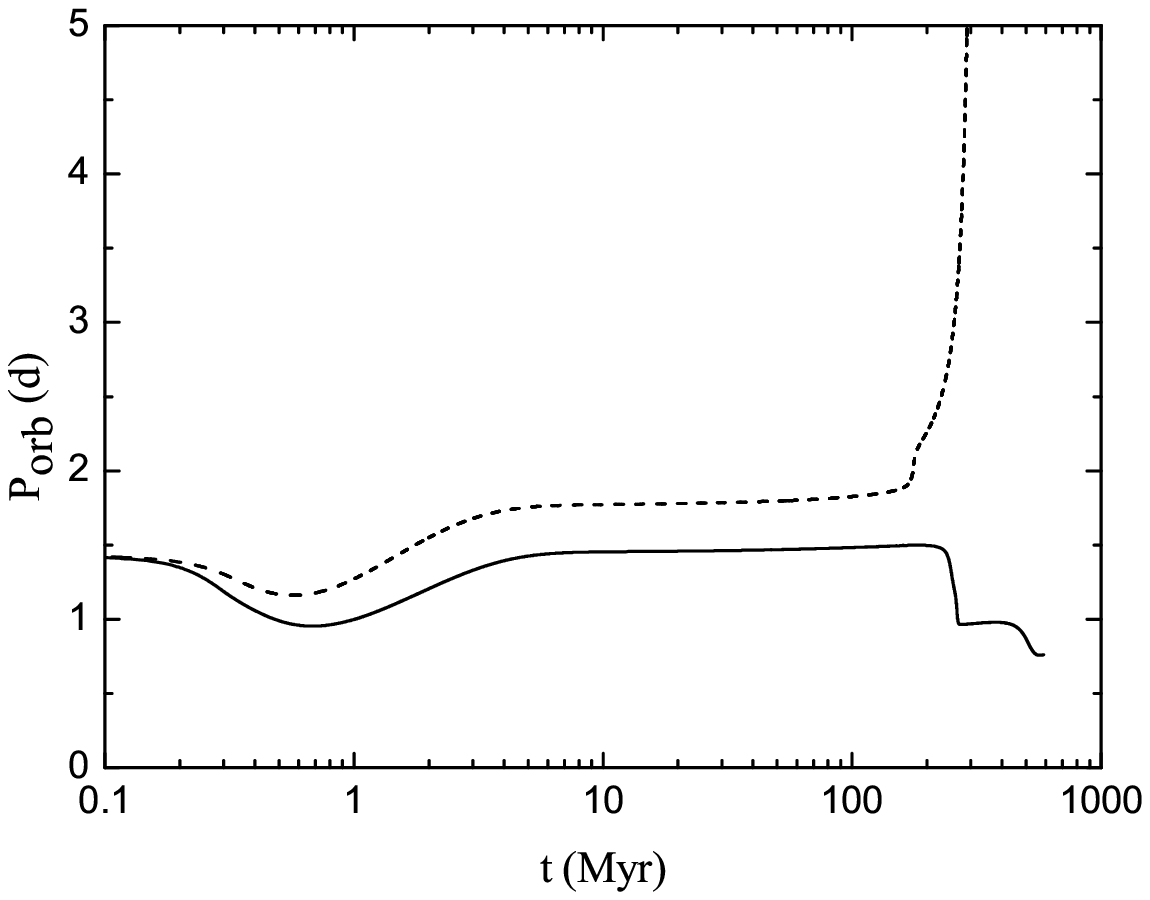} \plotone{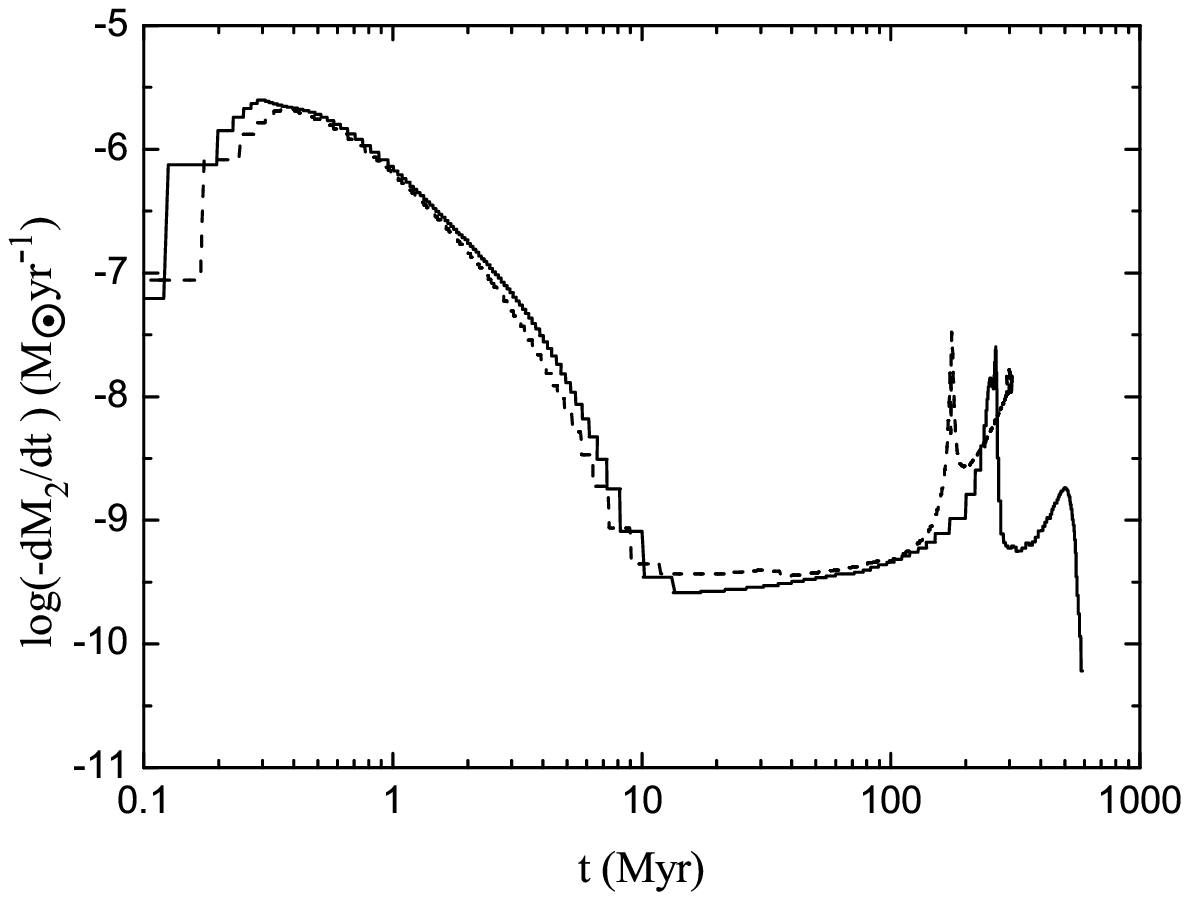}
\caption{The evolution of the mass ratio $q$ (left), the orbital
period $P_{\rm orb}$ (center), and the mass transfer rate
$\dot{M}$ (right) with the mass transfer time for a Algol binary
with the initial orbital period $P_{\rm orb,i}=1.4$ d. The solid
and dashed curves correspond to $\delta=0.01$ and 0.005,
respectively. The mass transfer rate have been slightly smoothed
for clarity.\label{fig2}}
\end{figure}

\noindent{\it (2) Initial masses $(M_2, M_1)=(3.0, 2.0) M_{\odot}$
and orbital period $P_{\rm orb,i}\simeq 3.4$ days}

The calculated results for a relative wide system with $P_{\rm
orb,i}\simeq 3.4$ days and $\delta=0.032$, 0.015 are presented in
Fig.~3. Compared with those in Fig.~2, the mass transfer rates are
much higher, and the evolutionary timescale is smaller by a factor
more than 100, suggesting that these systems are less likely to be
observed. For both values of $\delta$, the orbital period shows a
transition from decrease to increase when $q\sim 0.2-0.4$. The
phases of orbit shrinkage last about $20\%$ of the total mass
transfer lifetime. When $\delta=0.015$, the mass transfer rate
somewhat fluctuates during the late stages of the evolution though
we have smoothed the mass transfer rate for clarity.

\begin{figure}
\epsscale{.3} \plotone{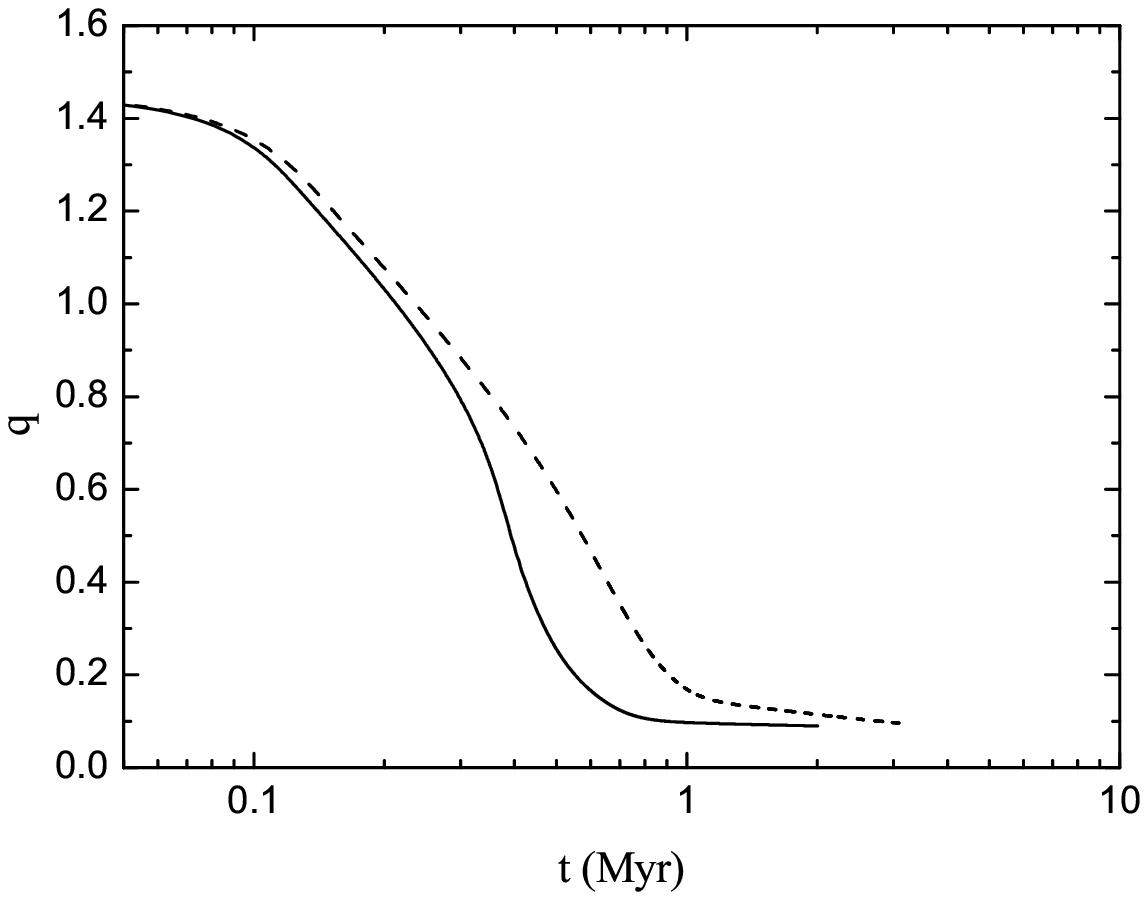}\plotone{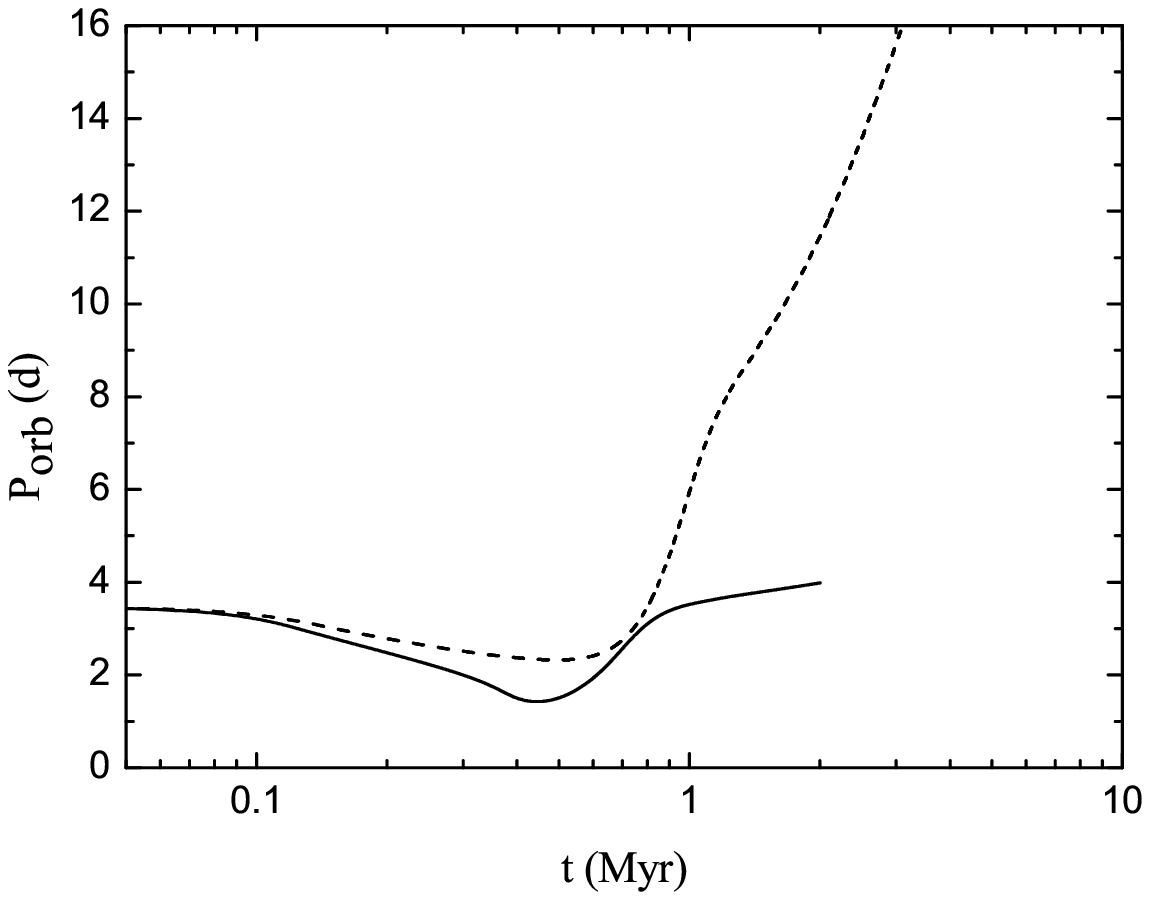}
\plotone{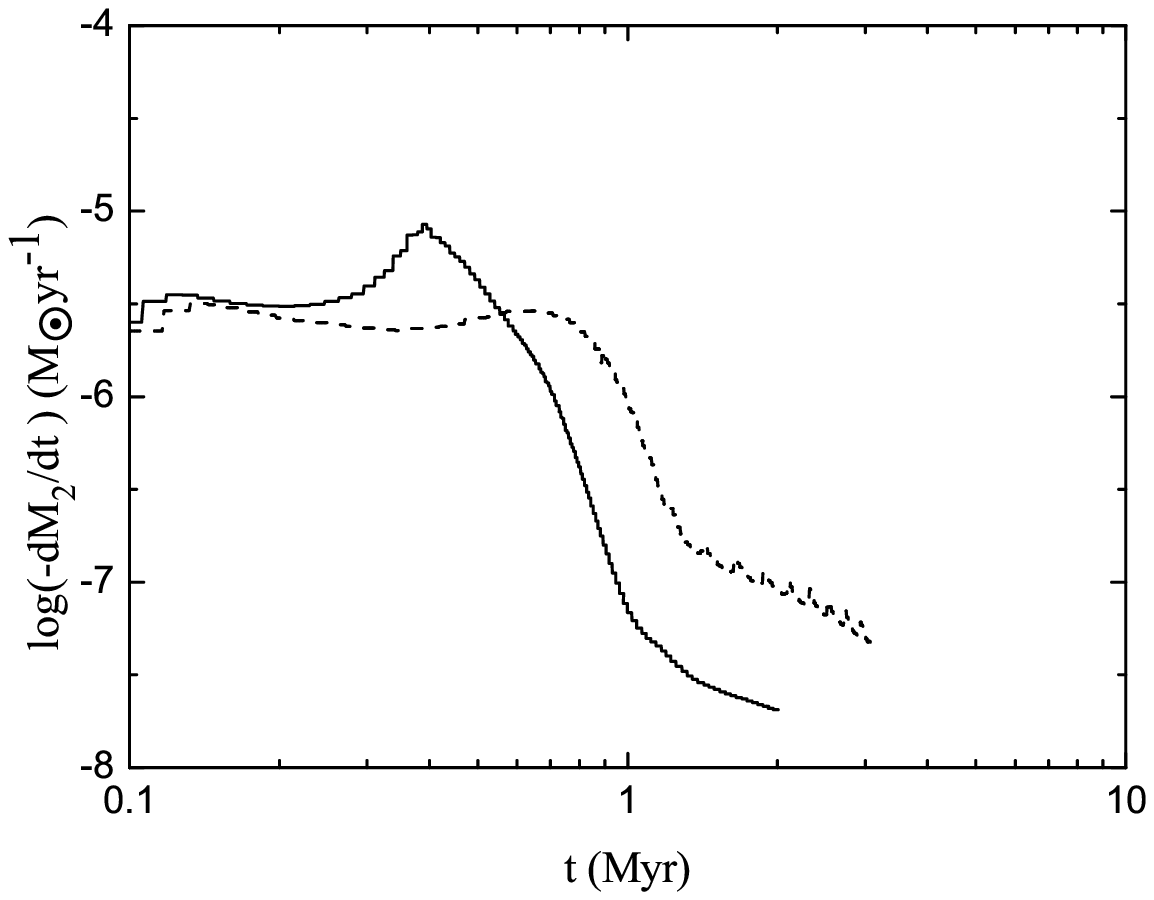}\caption{Same as Fig. 2, but for the initial
orbital period $P_{\rm orb,i}=3.4$ d. The solid and dashed curves
correspond to $\delta=0.032$ and 0.015, respectively. The mass
transfer rate have been slightly smoothed for
clarity.\label{fig3}}
\end{figure}

The calculated evolutionary tracks in the HR diagram are plotted
in Fig.~4. The bold curves denote the phases in which the orbital
period decreases. For comparison with observations, we also show
the locations of four Algol systems which have decreasing orbital
periods and well-measured parameters summarized in Table 1. As
shown in the figure, the calculated results are roughly in
accordance with the observations, suggesting that they may
represent the possible evolutionary sequences that lead to the
formation of these systems. Figure 5 compares the calculated
results of Algol  binary evolution with the observational data in
the $T_{\rm eff} - P_{\rm orb}$ plane. It seems that our model is
also compatible with the observed spectral types of the four
systems.

\begin{figure}
\epsscale{.95} \plotone{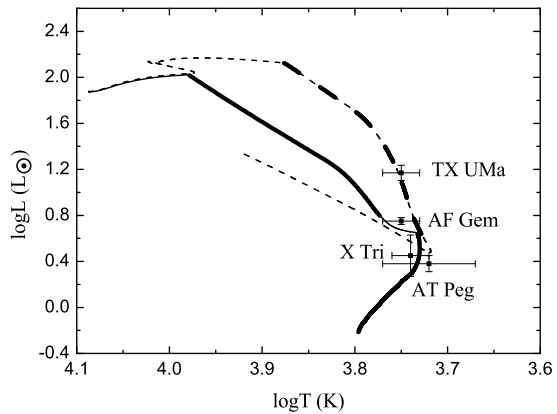} \caption{The calculated
evolutionary tracks for Algol binaries in the HR diagram. The
solid and dashed curves correspond to the initial orbital period
$P_{\rm orb,i}=1.4$ d $(\delta=0.01)$ and $P_{\rm orb,i}=3.4$ d
$(\delta=0.032)$, respectively. The bold curves denote the phases
in which the orbital period decreases. The filled rectangles
represent the locations of the four Algol binaries listed in Table
1. \label{fig4}}
\end{figure}

\begin{figure}
\epsscale{.95} \plotone{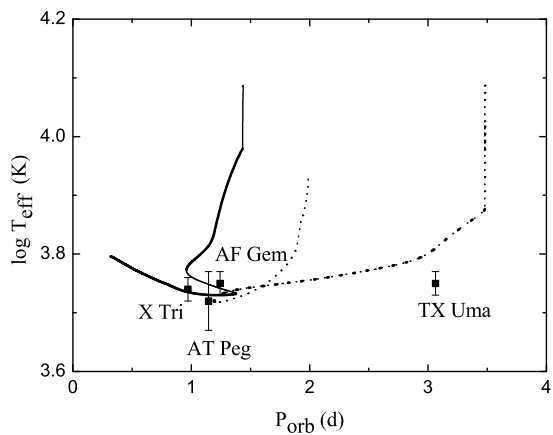} \caption{Evolutionary sequences in
the donor star's effective temperature $T_{\rm eff}$ versus
orbital period $P_{\rm orb}$ diagram. The solid and dotted curves
correspond to the initial orbital period $P_{\rm orb,i}=1.4$ d
$(\delta=0.01)$ and $P_{\rm orb,i}=3.4$ d $ (\delta=0.032)$,
respectively. The bold curves denote the phases in which the
orbital period decreases. The filled rectangles represent the
locations of the four Algol binaries listed in Table 1.
\label{fig5}}
\end{figure}



\section{Discussion and Conclusions}
It is an open question why some of the Algol systems show orbital
shrinkage while others not. Although it is generally believed that
there must be orbital angular momentum loss during the evolution
of these binaries, the appropriate mechanism(s) for angular
momentum loss has not been verified. Based on theoretical model of
\citet{spru01} and \citet{taam01}, we suggest that a CB disk may
play an important role in determining the orbital change in Algol
systems. Note that our CB disk hypothesis does not exclude the
effect of (magnetized) mass loss. On the contrary, the formation
of the CB disk may be closely related to the mass loss processes
in the binary evolution. The existence of the CB disks may be
revealed by their infrared radiation, as in GG Tau \citep{rodd96}.
The Monte-Carlo simulation by \citet{wood99} has reproduced the
morphology of the near IR radiation which is in accord with the
observations by \citet{rodd96}.

In this paper we have calculated the evolutionary sequence for
Algol binaries with typical masses and orbital periods, taking
into account both magnetic braking and the CB disk. The detailed
calculations indicate that the orbital evolution of Algol binaries
can be significantly affected by the CB disk, especially for Case
A mass transfer. With adequate values of $\delta$, it is possible
to account for the decrease of the orbital periods in some Algol
binaries. In addition, the existence of CB disk can accelerate the
evolution and enhance the mass transfer rates, as already noticed
in the evolutionary study of CVs \citep{taam01}. This suggests a
plausible explanation for the rapid mass transfer observed in a
few Algol binaries with very low mass ratios \citep{qian02a}.

Finally, we note that the magnitude of the mass feeding parameter
$\delta$ is highly uncertain, let alone its variation with the
mass transfer rates. These uncertainties make it difficult to
present direct comparison between observations and theoretical
predications for Algol binaries. However, our analysis provides a
reasonable mechanism for angular momentum loss in semi-detached
binaries including Algol binaries. More detailed multi-wavelength
observations can provide stringent tests for the CB disk model,
and the theories of the evolution of Algol binaries.

\acknowledgments {We thank the anonymous referee for his/her
helpful comments improving this manuscript. This work was
supported by Natural Science Foundation of China under grant
number 10573010.}

\begin{deluxetable}{ccccccccc}
\tabletypesize{\scriptsize} \rotate \tablecaption{Observed
Parameters of four Algol System \label{tbl-1}} \tablewidth{0pt}
\tablehead{ \colhead{Source} & \colhead{$P$ (d)} &
\colhead{$\dot{P}$($10^{-6}$ d\,yr$^{-1}$)} & \colhead{Spectrum} &
\colhead{$M_{1}(M_{\odot})$} & \colhead{$M_{2}(M_{\odot})$} &
\colhead{$\log L_{1}(L_{\odot})$} & \colhead{$\log
L_{2}(L_{\odot})$} & \colhead{References} } \startdata
X Tri  &0.9715397 &-0.142   &A2/G3  &2.3$\pm$0.7     &1.2$\pm$0.3       &1.21$\pm$0.06  &0.45$\pm$0.18  &1 \\
AT Peg &1.146082  &-0.638   &A4/G   &2.22$\pm$0.065  &1.05$\pm$0.025    &1.19$\pm$0.025 &0.38$\pm$0.07  &2 \\
AF Gem &1.2435012 &-0.0992  &B9/G0  &3.37$\pm$0.11   &1.155$\pm$0.038   &1.78$\pm$0.09  &0.75$\pm$0.03  &3 \\
TX UMa &3.0632881 &-0.713   &B8/G0  &4.76$\pm$0.16   &1.18$\pm$0.04     &2.30$\pm$0.04  &1.17$\pm$0.065 &4 \\
\enddata


\tablenotetext{a}{References: 1 - \citet{mez80}, 2 -
\citet{max94}, 3 - \citet{max95a}, 4 - \cite{max95b}.}

\end{deluxetable}


\begin{thebibliography}{}
\bibitem[Andronov, Pinsonneault, \& Sills(2003)]{andr03} Andronov, N., Pinsonneault, M., \& Sills, A. 2003, ApJ, 582, 358
\bibitem[Artymowicz \& Lubow(1994)]{arty94} Artymowicz, P., \& Lubow, S. H. 1994, \apj, 421, 651
\bibitem[Belle et al. (2004)]{bell04} Belle, K. E., Sanghi, N., Howell, S. B., Holberg, J. B., \& Williams, P. T. 2004, \aj, 128, 448
\bibitem[Caughlan \& Fowler(1988)]{caug88} Caughlan, G. R., \& Fowler, W. A. 1988, ADNAT, 40, 284
\bibitem[Chen et al.(2006)]{chen06} Chen, W.-C, Li, X.-D., \& Wang, Z.-R. 2006, \pasj, 58, 153
\bibitem[De Loore \& van Rensbergen(2004)]{loor04} De Loore, C. \& van Rensbergen,W. 2004, \apss, 296, 353
\bibitem[Eggleton(1971)]{eggl71} Eggleton, P. P. 1971, \mnras, 151, 351
\bibitem[Eggleton(1972)]{eggl72} Eggleton, P. P. 1972, \mnras, 156, 361
\bibitem[Eggleton(2000)]{eggl00} Eggleton, P. P. 2000, \na, 44, 111
\bibitem[Giuricin \& Mardirossian(1981)]{Giur81} Giuricin, G., \& Mardirossian, F. 1981, \apjs, 46, 1
\bibitem[Han et al.(1994)]{han94} Han, Z., Podsiadlowski, P., \& Eggleton, P. P. 1994, \mnras, 270, 121
\bibitem[Huang(1963)]{huan63} Huang, S.-S. 1963, \apj, 138, 471
\bibitem[Iglesias et al.(1992)]{igle92} Iglesias, C. A., Rogers, F. J., \& Wilson, B.G. 1992, \apj, 397, 717
\bibitem[Itoh et al.(1989)]{itoh89} Itoh, N., Adachi, T., Nakagawa, M., Kohyama, Y., \& Munakata, H. 1989, \apj, 339, 354
\bibitem[Kim \& Demarque(1996)]{kim96} Kim, Y.-C., \& Demarque, P. 1996, \apj, 457, 340
\bibitem[Li (2004)]{li04} Li, X. -D. 2004, \apj, 616, L119
\bibitem[Lloyd \& Guilbault(2002)]{lloy02} Lloyd, G. \& Guilbault, P. 2002, The Observatory, 122, 85
\bibitem[Maxted et al.(1994)]{max94} Maxted, P. F. L., Hill, G., \& Hilditch, R. W. 1994, \aap, 285, 535
\bibitem[Maxted \& Hilditch(1995)]{max95a} Maxted, P. F. L., \& Hilditch, R. W. 1995, \aap, 301, 149
\bibitem[Maxted et al.(1995)]{max95b} Maxted, P. F. L., Hill, G., \& Hilditch, R. W. 1995, \aap, 301, 135
\bibitem[Mezzetti et al.(1980)]{mez80} Mezzetti, M., Cester, B., Giuricin, G., \& Mardirossian, F. 1980, A\&AS, 39, 265
\bibitem[Nelson \& Eggleton(2001)]{nel01} Nelson, C. A. \& Eggleton , P. P. 2001, ApJ, 552, 664
\bibitem[Paczy\'nski(1971)]{pacz71} Paczy\'nski, B. 1971, ARA\&A, 9, 183
\bibitem[Peters(2001)]{pete01} Peters, G. 2001, \apss, 264, 79
\bibitem[Plavec(1968)]{plav68} Plavec, M. 1968, Adv. Astr. Ap., 6, 201
\bibitem[Pols et al.(1995)]{pols95} Pols, O., Tout, C. A., Eggleton, P. P., \& Han, Z. 1995, \mnras, 274, 964
\bibitem[Qian(2000a)]{qian00a} Qian, S.-B. 2000a, \aj, 119, 901
\bibitem[Qian(2000b)]{qian00b} Qian, S.-B. 2000b, \aj, 119, 3064
\bibitem[Qian(2001a)]{qian01a} Qian, S.-B. 2001a, \aj, 121, 1614
\bibitem[Qian(2001b)]{qian01b} Qian, S.-B. 2001b, \aj, 122, 1561
\bibitem[Qian(2001c)]{qian01c} Qian, S.-B. 2001c, \aj, 122, 2686
\bibitem[Qian(2002)]{qian02} Qian, S.-B. 2002, \pasp, 114, 650
\bibitem[Qian et al.(2002)]{qian02a} Qian, S.-B. et al. 2002, \aap, 396, 609
\bibitem[Queloz(1998)]{quel98} Queloz, D., Allain, S., Mermilliod, J. C., Bouvier, J., \& Mayor, M. 1998, \aap, 335, 183
\bibitem[Rappaport et al.(1983)]{rapp83} Rappaport, S., Verbunt, F., \& Joss, P. C. 1983, \apj, 275, 713
\bibitem[Refsdal, Roth, \& Weigert(1969)]{refs74} Refsdal, S., Roth, M. L., \& Weigert, A., 1974, A\&A, 36, 113
\bibitem[Refsdal \& Weigert(1969)]{refs69} Refsdal, S. \& Weigert, A., 1969, A\&A, 1, 16
\bibitem[Roddier et al. (1996)]{rodd96} Roddier, C., Roddier, F., Northcott, M. J., Graves, J. E., \& Jim, K. 1996, \apj, 463, 326
\bibitem[Sarna(1993)]{sarn93} Sarna, M. 1993, MNRAS, 262, 534
\bibitem[Sarna, Muslimov, \& Yerli(1997)]{sarn97} Sarna, M., Muslimov, A., \& Yerli, S. K. 1997, MNRAS, 286, 209
\bibitem[Sarna, Yerli, \& Muslimov(1998)]{sarn98} Sarna, M., Yerli, S. K., \& Muslimov, A. 1998, MNRAS, 297, 760
\bibitem[Schatzman(1962)]{scha62} Schatzman, E. 1962, AnAp, 25, 18
\bibitem[Shakura \& Sunyaev(1973)]{Shak73} Shakura, N. I., \& Sunyaev, R. A. 1973, \aap, 24,  337
\bibitem[Sills et al.(2000)]{sill00} Sills, A., Pinsonneault, M. H., \& Terndrup, D. M. 2000, \apj, 534, 335
\bibitem[Spruit \& Cao(1994)]{spru94} Spruit, H. C., \& Cao, X. 1994, \aap, 287, 80
\bibitem[Spruit \& Taam(2001)]{spru01} Spruit, H. C., \& Taam, R. E. 2001, \apj, 548, 900
\bibitem[Taam \& Spruit(2001)]{taam01} Taam, R. E., \& Spruit, H. C.  2001, \apj, 561, 329
\bibitem[Thomas(1977)]{thom77} Thomas, M. C. 1977, ARA\&A, 15, 127
\bibitem[van den Heuvel \& de Loore(1973)]{van73} van den Heuvel, E.P.J., \& de Loore, C. 1973, \aap, 25, 387
\bibitem[van den Heuvel(1994)]{van94} van den Heuvel, E.P.J. 1994 in Interacting Binaries (Saas-Fee 22), Shore, S. N., et al., eds., p263
\bibitem[Verbunt \& Zwaan(1981)]{verb81} Verbunt, F., \& Zwaan, C. 1981, \aap, 100, L7
\bibitem[Warner(1987)]{warn87} Warner, B. 1987, \mnras, 227, 23
\bibitem[Wood et al. (1999)]{wood99} Wood, K., Crosas, M., \& Ghez, A. 1999, \apj, 516, 335
\end{thebibliography}
\end{document}